\documentstyle{article}
\begin{document}

\title{Steps to the Planck Function : A Centenary Reflection}
\author{T Richard Carson \\
School of Physics and Astronomy, University of St Andrews} 
\maketitle
\begin{abstract}
The year 2000 marks the centenary of the Planck function for the blackbody 
spectral distribution. Here are traced the steps, over 100 years,  
along the path leading to this important formula.
\end{abstract}
\section{Introduction}

The Planck function was first presented at a meeting of the German Physical 
Society in Berlin on 19 October 1900 in a contribution entitled "On an 
improvement of the Wien spectral equation", and published in the 
proceedings of the society in a brief three-page report (Planck, 1900Va). 
Only the functional form was given, including two undetermined constants,   
derived on the basis of a tentative relation between the entropy and the 
energy of each of a collection of radiation resonators.
The definitive formula was presented at a meeting of 
the society on 14 December 1900 in a paper entitled "On the theory of the law 
of the energy distribution in the normal spectrum" and published,   
in a nine-page report (Planck, 1900Vb).  
Herein Planck introduced his new natural constant $h$ relating the "energy 
element" $\epsilon $ associated with a resonator of frequency $\nu $ by the 
formula $\epsilon = h \nu $, with the numerical value $h = 6.55 x 10^{-27}$
erg seconds. He only sketches the actual calculation of the entropy,
now for the first time related to the probability of resonator 
complexions via the Boltzmann principle.
Thus ended the long search for the blackbody spectral 
distribution function and simultaneously began the era of quantum physics.
A full treatment of the derivation was later given in a paper (Planck, 1901a) 
entitled "On the law of the energy distribution in the normal spectrum" 
submitted to Annalen der Physik on 9 January 1901.

\section{Thermal Radiation}

That the arrival of both the visible light and the 'dark' radiant heat 
from the Sun are simultaneously interrupted and restored at the time of a solar 
eclipse, indicated that light and heat are propagated across space 
with the same velocity. The astronomer W. Herschel (1800), using a
thermometer exposed to the prismatic spectrum of the Sun, discovered that the
heating properties of light extended beyond the visible spectrum,
being enhanced most especially in the infrared. He
also showed that the unseen heat radiation obeyed the same laws of 
reflection and refraction as visible light. The similarity in
other properties, including bi-refringence and polarisation, as found by
others, for light and heat, illustrated a difference of degree rather
than kind. Furthermore, that sources of light are also sources of heat, 
and ${\em vice\; versa}$, served to distinguish the two manifestations 
only by the means of detection employed.
Indeed, from his measurements, Herschel presented two distributions of the 
(continuous) intensity of the solar rays, based respectively on their ability 
to illuminate and to heat. Because of the two different techniques of 
measurement, the relative scales are not related. If the two intensity 
distributions were brought to a common scale and then summed, it would 
represent the first wideband total intensity distribution for the solar spectrum.

Simple experiments, such as those by Leslie (1804) and others, showed that  
the rate of emission of thermal radiation from a heated body, expressed as
the emissive power or radiant emittance E, being the power emitted per unit 
area, depends on the temperature and on the nature of the surface. Thus a
blackened surface has a higher emittance than a polished one at the same
temperature. Also when two such surfaces are subjected to incident radiant
energy the fraction of energy absorbed, or absorptivity $A$, depends on the
temperature and on the nature of the surface, a blackened surface being a more
efficient absorber than a polished one.

Kirchhoff (1859), by means of a thought experiment, demonstrated
 that the ratio $E/A$ is the same for all bodies at the same
temperature, emitting and absorbing at the same wavelength. He considered
two bodies, in the form of infinte surfaces, S$_{1}$ and S$_{2}$, backed 
with perfectly reflecting mirrors and arranged parallel to each other with
the mirrors on the outside. S$_{1}$ is such that it can only emit and absorb
radiation of wavelength $\Lambda $, while S$_{2}$ can emit and absorb all
possible wavelengths $\lambda $. At wavelength $\Lambda $ the emissivities
are $E_{1}$ and $E_{2}$, and the absorptivities are $A_{1}$ and $A_{2}$.
For all $\lambda \neq \Lambda $ all the radiation emitted by S$_{2}$ is
reabsorbed by it. Therefore for temperature equilibrium S$_{2}$ must absorb
as much as it emits of the radiation of wavelength $\lambda = \Lambda $,
whether it comes from S$_{1}$ or S$_{2}$. Of the radiation $E_{1}$ emitted
by S$_{1}$, S$_{2}$ absorbs, after repeated absorptions and emissions :
\[A_{2}E_{1} \sum _{n=0}^{\infty } k^{n} = A_{2}E_{1}/(1-k) \]
while of the radiation $E_{2}$ emitted by S$_{2}$ itself, S$_{2}$ absorbs:
\[(1-A_{1})A_{2}E_{2} \sum _{n=0}^{\infty } k^{n} = (1-A_{1})A_{2}E_{2}/(1-k) \]
where $k = (1 - A_{1})(1 - A_{2})$. Hence :
\[E_{2} = A_{2} [E_{1} + (1 - A_{1}) E_{2}]/(1 - k) \]
giving :
\[ A_{1} E_{2} = A_{2} E_{1} \;\; {\rm or} \;\; E_{1}/A_{1} = E_{2}/A_{2} \]
The same result is obtained on applying the condition of temperature
equilibrium to S$_{1}$, yielding :
\[E_{1} = A_{1} [E_{2} + (1 - A_{2}) E_{1}]/(1 - k) \]
Since the composition of the surfaces, as well as the wavelength $\Lambda $
and the temperature are arbitrary, it follows that the ratio $E/A$,
at any wavelength or temperature, is the same for 
 all surfaces, and is a function only of wavelength and temperature,
which statement is known as Kirchhoff's Law.

Kirchhoff (1860) generalised the above argument, which involved only discrete
wavelengths, to the continuous spectrum, by re-defining $E$ as the emittance
per unit wavelength interval. He also introduced the concept of a 
{\em perfectly black body}, being one which absorbs all incident radiation, for
 which $A = 1$ at all wavelengths and temperatures. By considering the temperature
equilibrium of a black body enclosed in a cavity with walls made of a similarly
black material, he showed that the emittance is equal to the radiant power, 
$F({\lambda },T)$ per unit wavelength interval, crossing unit area, within
the `black body' enclosure. Hence, by considering the temperature equilibrium
of a {\em non-black} body within the cavity, Kirchhoff's Law takes the form 
$E/A = F(\lambda , T)$.
Thus the absorptivity of a body is equal to its relative emittance,
or the emittance of the body relative to that of a black body.
Kirchhoff stressed the importance of the study of black body radiation
in order to determine the function $F(\lambda ,T)$.
It is also worth noting here that the enunciation by Kirchhoff of the 
principle of detailed balance between emission and absorption precedes that 
of Einstein by more than half a century. 
\section{Integrated Properties of Black Body Radiation}

Given the dependence on wavelength $\lambda $ and temperature $T$
of the black body radiation intensity $I(\lambda , T)$,
its energy density $u$ and flux $F$, both monochromatic and integrated, can be
obtained :
\[I(T) = \int I(\lambda , T){\rm d}\lambda \]
\[u(\lambda ,T) = (4\pi /c)I(\lambda ,T),
  u(T) = \int u(\lambda , T){\rm d}\lambda  = (4\pi/c)I(T) \]
\[F(\lambda , T) = \pi I(\lambda , T),
  F(T) = \int F(\lambda , T){\rm d}\lambda   = (c/4) u(T) \]

As indicated by Kirchhoff, the properties of black body radiation can be studied
by observations on the radiation emerging from a small orifice in the wall
of a black body cavity. Experiments by Stefan (1879) established that the
total (integrated) emergent flux could be represented in the form :
\[F(T) = \sigma T^{4} \]
where $\sigma  = 5.670 x 10^{-8} J m^{-2} s^{-1} K^{-4}$ is the 
Stefan(-Boltzmann) constant, from which it follows that the energy density
is given by :
\[u(T) = (4\sigma /c) T^{4} = a T^{4} \]
with the radiation constant $a = 7.565 x 10^{-16} J m^{-3} K^{-4}$. The
latter form  was established theoretically by Boltzmann (1884) on the basis
of thermodynamics. Using for the pressure of radiation the relation :
\[P(T) = (1/3)u(T) \]
as indicated by the electromagnetic theory of Maxwell (1873), and assuming
that radiation obeyed the known laws of thermodynamics, then putting $U = uV$ :
\[{\rm d}Q = T{\rm d}S = {\rm d}(uV) + P {\rm d}V = (u + P){\rm d}V + V {\rm d}u \]
from which it follows that :
\[(\partial /\partial u)_{V}[(u + P)/T] = (\partial /\partial V)_{u}[V/T] \]
which, on using $P = u/3$, gives :
\[(T/u)(\partial u/\partial T)_{V} = (\partial \ln u/\partial \ln T)_{V} = 4 \]
so that :
\[u = aT^{4}, \;\;\; P = (1/3)aT^{4} \]
in agreement with the result of Stefan. Alternatively, Stefan's result
may be taken as a proof of the relation $P = u/3$. 
From the above expressions for $u$ and $P$ it follows that :
\[{\rm d}S = (4/3)aT^{3}{\rm d}V + 4aT^{2}V{\rm d}T \]
giving for the entropy :
\[S = (4/3)aT^{3}V \]

\section{Spectrum of Black Body Radiation}

The challenge, first presented by Kirchhoff (1859), to determine the spectral
distribution of black body radiation, was not to be completely answered
until the turn of the century. Experiments by Draper (1847) 
and by Weber (1887) showed that,
with increasing temperature, the maximum intensity in the light emitted
by incandescent solid bodies, progressed across the spectrum, in the
direction from the red to the violet. Bartoli (1876) had pointed out that,
in order not to contradict the second law of thermodynamics, radiation
must exert a pressure, and Boltzmann (1884) showed that work must be
done against this pressure in compressing a volume occupied by radiation alone.
Wien (1893), developing this argument, made the first important steps
in the determination of the black body radiation spectrum.
On the basis of the Doppler principle, applied to radiation within
a rectangular parallelepiped and reflected (normally) from a
wall moving with velocity $v$, where the relative wavelength
change is given by 
$\delta \lambda /\lambda = 2 v/c$
(twice the ordinary Doppler shift), then radiation of wavelength $\lambda _{0}$
and frequency $\nu _{0}$ in a volume $V_{0}$, subjected to an adiabatic
compression to volume $V$, undergoes a change to wavelength $\lambda $ and
frequency $\nu $ where :
\[\lambda ^{3} V^{-1} = \lambda _{0}^{3} V_{0}^{-1} \;\;\;{\rm and}\;\;\;
  \nu ^{3} V = \nu _{0}^{3} V_{0} \]
The change in energy of the radiation, supplied by the work done
in the compression against the radiation pressure,
results in a change in temperature from $T_{0}$ to $T$.
Since for an adiabatic change 
$VT^{3} = V_{0}T_{0}^{3}$ it follows that :
\[\lambda T = \lambda _{0}T_{0}\;\;\; {\rm and}\;\;\; \nu /T = \nu_{0}/T_{0} \]
giving the earliest expression of the Wien Displacement Law, which states
that  with changing temperature the black body spectrum is shifted
so that for each wavelength the product of the temperature and the
wavelength remains constant. Similarly each interval of wavelength and
frequency is changed according to :
\[T {\rm d}\lambda  = T_{0}{\rm d}\lambda _{0} \;\;\; {\rm and} \;\;\;
  {\rm d}\nu /T = {\rm d}\nu _{0}/T_{0} \]
Since the (spatial) energy density obeys the relation :
\[u(\lambda ,T){\rm d}\lambda /u(\lambda _{0},T_{0}){\rm d}\lambda _{0} = 
  u(\nu ,T){\rm d}\nu /u(\nu _{0},T_{0}){\rm d}\nu _{0} = T^{4}/T_{0}^{4} \]
then :
\[u(\lambda ,T)/u(\lambda _{0},T_{0}) = (T/T_{0})^{5} = (\lambda _{0}/\lambda )^{5} \]
\[u(\nu ,T)/u(\nu _{0},T_{0}) = (T/T_{0})^{3} = (\nu /\nu _{0})^{3} \]
in agreement with the observations of Weber. It also follows that the
wavelength $\lambda _{m}(T)$ and frequency $\nu _{m}(T)$ at which
the wavelength and frequency spectral distributions reach their maxima
satisfy the relations :
\[T \lambda _{m}(T) = T_{0}\lambda _{m}(T_{0}) \;\;\; {\rm and} \;\;\;
  \nu _{m}(T)/T = \nu _{m}(T_{0})/T_{0} \]
representing another expression of the Wien Displacement Law.
Therefore given the spectral distribution for one temperature $T_{0}$,
that for any other temperature $T$ can be determined.

Wien (1894) re-derived all the above relations when the radiation is
enclosed in a spherical volume of variable radius. He also showed that the  
spectral distributions for different temperatures cannot cross. The condition 
for this to be true is that :
\[[\partial \ln u(\lambda ,T)/\partial \ln \lambda ]_{T} > -5 \] 
which is trivially true for the rising branch of the curve, and also
obviously true for the falling branch. In terms of the frequency
distribution the condition becomes :
\[[\partial \ln u(\nu ,T)/\partial \ln \nu ]_{T} < + 3 \]

\subsection{The Wien Formula}

Wien (1896) pointed out that, since the black body radiation spectrum
corresponded to thermal equilibrium, it should be completely determined
from the principle of maximum entropy, although he did not do this
himself. Instead, adopting a radiation model based on  analogy with
the Maxwell velocity distribution, he obtained a spectrum of the form :
\[u(\lambda ,T) = f_{1}(\lambda )\exp[-f_{2}(\lambda )/T] \]
where $f_{1}(\lambda ) = c_{1}/\lambda ^{\alpha }$, ($c_{1}$ and $\alpha $
being constants), and $f_{2}(\lambda )$ are functions to be determined.
In order to satisfy the Stefan-Boltzmann $T^{4}$ dependence of the
energy density, it is necessary for $\alpha = 5$ and for $f_{2}(\lambda )/T$
to be a function of the product $\lambda T$, thus giving a more general
mathematical expression to the Wien Displacement Law. Putting, specifically,
$f_{2}(\lambda )/T = c_{2}/(\lambda T)$, where $c_{2}$ is a constant,
gives the form :
\[u(\lambda ,T) = c_{1}\lambda ^{-5} \exp [-c_{2}/(\lambda T)] \]
known as Wien's formula, which attains its maximum value
at a wavelength $\lambda _{m}$ with $\lambda _{m}T = c_{2}/5$.
The formula was in good accord with the observations of Paschen (1896), with
$c_{2} = 5\lambda _{m}T = 14455 \mu \;$K, or $\lambda_{m}T = 2.891x10^{-3}$ m K.

In a series of studies, from 1897 onwards, on the entropy of radiation, 
treating the radiation field as an ensemble of monochromatic resonators,
Planck sought to realize the hope of Wien that the blackbody radiation 
spectrum could be derived on the basis of the maximization of the entropy. 
Indeed Planck (1899) believed (erroneously) that he had shown that the 
Wien spectral distribution was a necessary consequence of the principle of
entropy maximization, and that its range of validity coincided with that of 
the (two) laws of thermodynamics. Such support for the Wien formula led to 
it being referred to as the Wien-Planck distribution.
However, experimental measurements by Lummer and Pringsheim (1899), 
Rubens (1899) and others pointed to systematic departures of the Wien-Planck  
formula from the experimental results particularly at long wavelengths.
Thiesen (1900), following the lead of Wien, expressed the spectral
distribution in the, only slightly different, general form :
\[u(\lambda ,T) = T^{5} \psi (\lambda T) \]
where $\psi (\lambda T)$ is a still unknown function.
Thiesen also suggested that the function $\psi $, when determined, would 
accommodate the deviations between theory and experiment by 
differing from the Wien-Planck form, leading to a modification of the 
Wien displacement law.

Planck advanced his support of the Wien formula in two further 
papers (Planck, 1900Aa and 1900Ab).  In the first paper he adopted, by way of 
definition, $s = -(w/a\nu )\ln (w/eb\nu )$, where $s$ and $w$ are the 
entropy and (average) energy per radiation resonator of frequency $\nu $, 
and $a$ and $b$ are universal positive constants, while $e$ is the base of 
natural (Napierian) logarithms. Using the thermodynamic relation 
$ds/dw = 1/T$ where $T$ is the temperature, it follows that 
$w = b\nu \exp (-a\nu /T)$, from which, on multiplying by the density 
$8\pi \nu ^{2}/c^{3}$ of resonators in $\nu $-space, the Wien formula follows. 
From the measurements of Paschen (1899) the constants $a$ and $b$ were 
determined to be $a = 4.818 x 10^{-11}$ sec K and $b = 6.885 x 10{-27}$ 
erg sec. The value of $b$ here therefore represents the first determination 
of what later became the Planck constant $h$, while the value of $a$ is that 
of $h/k$.
In the second paper Planck sought to justify his earlier assumed definition 
of the entropy $s$, by arguing that maximization required 
$d^{2}s/dw^{2} = - F(w)$, where $F(w)$ is a positive function of $w$. 
Using a further relation (of limited validity) $F(nw) = F(w)/n$, where $n$ 
is the number of resonators, Planck adopted the solution $F(w) \propto 1/w$.
Thence putting $d^{2}s/dw^{2} = - \alpha /w$, it follows that 
$s = -\alpha w \ln (\beta w/e)$, where now $\alpha $ and $\beta $ are positive 
constants which depend only on the frequency $\nu $.  Then 
$w = [1/e\beta (\nu )] \exp  [-1/T\alpha (\nu)]$, leading again to a spectral 
distribution of the Wien form. Furthermore the Wien displacement law 
required both $\alpha $ and $\beta $ to be proportional to $\nu ^{-1}$, thus 
permitting the identifications $\alpha = 1/a\nu $ and $\beta = 1/eb\nu $,
where $a$ and $b$ are the quantities introduced earlier. However further 
work by Lummer and Pringsheim (1900) showed that both $a$ and $b$ were not 
constant but increased with wavelength, leading them to conclude the 
invalidity of the Wien-Planck spectral distribution.

\subsection{The Rayleigh Formula}

To Rayleigh (1900) the Wien formula  was difficult to accept on the grounds
it implied that, with increasing temperature, the radiation density at a given
wavelength approached a limit. Using an argument based on the still
disputed Maxwell-Boltzmann principle of the equipartition of energy,
in which each degree of freedom has an energy proportional to the
temperature $T$, and in analogy with the theory of sound, taking for
the density of vibrational modes $n(\lambda ) \propto \lambda ^{-4}$,
he obtained :
\[u(\lambda ,T) = K T \lambda ^{-4} \]
where $K$ is a constant, which has the form prescribed by Thiesen,
suggesting that it may be correct for large values of $\lambda T$.
However the formula tends to infinity at short wavelengths, leading
Rayleigh to attach the Wien exponential factor $\exp [- c_{2}/(\lambda T)]$.

\subsection{The Planck Formula}
In the light of the increasing evidence for the inadequacy of the Wien 
blackbody spectral distribution, Planck (1900Va) now addressed the problem 
of improving upon it. He recalled his earlier conclusion that the maximization  
of entropy was not of itself sufficient to derive the correct distribution,  
and recognised that his support of the Wien function had depended upon his 
adoption of the further special condition $d^{2}s/dw^{2} = - \alpha /w$, 
applicable only in the case of indefinitely small values of $w$ and $s$. 
Accordingly he considered instead the condition $d^{2}s/dw^{2} = - \alpha / w(
\beta + w)$, which he described as by far the next simplest relation giving 
$s$ as a logarithmic function of $w$, while at the same time reducing to
the former relation for vanishingly small values of $w$. It is readily shown 
that, ignoring constants of integration, $ds/dw = - (\alpha /\beta )
\ln \{w/(\beta + w)\}$, and $s = -(\alpha /\beta )
\{w \ln w - (\beta + w) \ln (\beta + w)\}$. Then, using $ds/dw = 1/T$, it 
follows that $w = \beta /\{\exp (\beta /\alpha T) -1\}$.
Noting that the Wien Displacement Law requires (see later) that $s$ be a 
function of $w/\nu $, where $\nu $ is the frequency, (demanding that 
$\beta $ be proportional to $\nu $), then enabled Planck to express the 
spectral distribution in the form 
$u(\lambda ,T) = C_{1} \lambda ^{-5}/\{\exp (C_{2}/\lambda T) - 1\}$, where  
$C_{1}$ and $C_{2}$ are constants, thus introducing what is now known as the 
Planck Function. Rubens and Kurlbaum (1900) pointed out that the new Planck 
formula included those of Wien and Rayleigh in the limiting cases of 
short and long wavelengths respectively, while their experimental 
measurements favoured it over all others.

However, the derivation was largely phenomenological since the entropy had 
not been calculated from first principles. Therefore Planck (1900Vb) next 
sought to follow the "Boltzmann prescription" whereby entropy involves 
disorder through the relation $S = k \ln W$, where $W$ is the number of ways 
or states in which the physical system may be realized.  
This use of this famous relation, never given by Boltzmann himself yet 
inscribed on his tombstone, is the first occurrence known to the writer.
Since the total 
radiation energy is distributed among all the radiation resonators 
involving a number of frequencies, it is first necessary to consider the 
distribution of an energy $E$ among the $N$ resonators with the same 
frequency $\nu $. Now if $E$ were infinitely divisible the distribution 
would be possible in infinitely many ways. Hence Planck considered 
- and this is the essential point of the whole calculation - $E$ to be made 
up of a fixed number $P$ of finite equal parts, or energy elements, each of 
magnitude $\epsilon = h \nu $, where $h$ is a new constant of nature, 
so that $E = P\epsilon = Nw$, where $w$ is the average energy per resonator.
This therefore represents the first appearance of the Planck constant, both 
as to its symbol and definition. Planck also proceeded to use combination 
theory to calculate the total number of possible complexions :
\[W = (N + P -1)!/(N-1)!/P! \approx (N + P)^{N+P}/N^{N}/P^{P} \]
on making use of the Stirling formula whereby for large $n, n! \approx n^{n}$.
He did not however give an expression for the entropy $S$, but gave his final 
result for the spectral distribution in its definitive form :
\[u(\nu ,T) = (8\pi h \nu ^{3}/c^{3})/[\exp (h \nu /kT) - 1] \]
where now the Boltzmann constant $k$ appears explicitly as well as the new 
(Planck) constant $h$. From the formula, using the measurements of 
Kurlbaum (1898) and of Lummer and Pringsheim (1900), the values of $h$ and 
$k$ were determined to be $h = 6.55 x 10^{-27}$ erg sec and 
$k = 1.346 x 10^{-16}$ erg grad. 

Very shortly afterwords, Planck (1901a) completed his derivation with details of the calculation of the entropy giving :
\[\ln W = N[(1+P/N)\ln (1+P/N) - (P/N) \ln(P/N)] \]
so that, since $P/N = w/\epsilon$, the entropy per resonator becomes :
\[s = S/N = k[(1 + w/\epsilon )\ln (1 + w/\epsilon) - 
(w/\epsilon )\ln (w/\epsilon)] = f(w/\epsilon ) \]
which for small values of $w/\epsilon $ (small $w$ and large $\epsilon $ or 
$\nu $) leads to the earlier (Wien) result. Now it follows that :
\[{\rm d}s/{\rm d}w = 1/T = (k/\epsilon )\ln (1 + \epsilon /w) \]
giving :
\[w = \epsilon /\{\exp (\epsilon /kT) - 1\} \]
Then since :
\[u(\nu ,T) = 8\pi (\nu ^{2}/c^{3}) w \]
and, according to the Wien Displacement Law :
\[u(\nu ,T) = \nu ^{3} f_{\nu }(\nu /T) \]
where $f_{\nu }$ is a still undetermined function :
\[w = \nu [(c^{3}/8\pi ) f_{\nu }(\nu /T)] \]
Thus :
\[1/T = {\rm d}s/{\rm d}w = \nu ^{-1} (8\pi/c^{3}) f_{\nu }^{-1}(w/\nu ) \]
so that, on integrating :
\[s = F(w/\nu ) \]
and therefore, in order to agree with the result that $s = f(w/\epsilon )$,
it follows that $\epsilon =h\nu $ giving :
\[w = h\nu /[\exp (h\nu /kT) - 1] = (hc/\lambda )/[\exp (hc/kT\lambda ) - 1] \]
where $h$ is again the Planck constant. Thus one finally 
obtains the Planck formula :
\[u(\nu ,T) = (8\pi h\nu ^{3}/c^{3})/[\exp (h\nu /kT) - 1] \]
\[u(\lambda ,T) = (8\pi hc/\lambda ^{5})/[\exp (hc/kT\lambda ) - 1] \]

The integrated radiation energy density is now $u(T) = a T^{4}$ with 
(Planck, 1901a) :
\[a = (8\pi k^{4}/h^{3}c^{3}) \int _{0}^{\infty }z^{3}/(\exp z - 1){\rm d}z = 
8\pi ^{5}k^{4}/15c^{3}h^{3} \]
in which the integral evaluates to $3! \zeta (4) = \pi ^{4}/15$ giving 
 $a = 7.565 x 10^{-16} m^{-3} K^{-4}$.
From the expression for w, giving 
$s = k[z/(\exp z -1) + \ln \{\exp z /(\exp z -1)\}]$, where $z = h\nu /kT$, 
the integrated entropy density is $S(T) = (4/3)aT^{3}V$, as has already been
derived from thermodynamics.

The maximum of the wavelength distribution occurs for 
$x_{m} = hc/kT\lambda _{m}$ where :
\[x_{m} = 5[1 - \exp (-x_{m})] = 4.965114...                \]
while the maximum of the frequency distribution occurs for 
$y_{m} = h\nu _{m}/kT$ where :
\[y_{m} = 3[1 - \exp (-y_{m})] = 2.821439...                \]
giving $\nu _{m}\lambda _{m}/c = y_{m}/x_{m} = 0.568252...  $, indicating
that $\lambda _{m} < c/\nu _{m}$ and $\nu _{m} < c/\lambda _{m}$.
Hence also :
\[\nu _{m}/T = (k/h)x_{m} = 5.8785 x 10^{10} Hz K^{-1} \]
\[\lambda _{m}T = (hc/k)/y_{m} = 2.89779 x 10^{-3} m K \]
In the low frequency/long wavelength limit $h\nu /kT = hc/kT\lambda  
\ll 1$ the Rayleigh formula is obtained, on expanding the exponential in the
denominator :
\[u(\nu ,T) = (8\pi h\nu ^{2}/c^{3})kT \]
\[u(\lambda ,T) = 8\pi kT/\lambda ^{4} \]
while in the high frequency/short wavelength limit $h\nu /kT = hc/kT\lambda \gg 1$
the Wien formula is obtained, on ignoring the 1 in the denominator :
\[u(\nu ,T) = (8\pi h\nu^{3}/c^{3}) \exp (-h\nu /kT) \]
\[u(\lambda ,T) = (8\pi hc/\lambda ^{5}) \exp (-hc/kT\lambda ) \]

Further experimental measurements by Lummer and Pringsheim (1901) 
showed that the values of $\lambda _{m} T$ were in excellent agreement 
with the predictions of Planck. Finally, Planck (1901b) returned to the 
question of the entropy of radiation, to demonstrate that the expression 
obtained from his theory satisfied the maximization principle. 
Thus in his stumbling search for the blackbody radiation spectral distribution,
Planck, in forging the hitherto missing link between radiation and 
thermodynamics, not only closed the era of classical physics but also 
opened up the era of quantum physics.

\section{Biographical Note}
Max (Karl Ernst Ludwig) Planck was born on 23 April 1858 in Kiel, and died 
on 4 October 1947 in G\"ottingen. He attended the universities of Munich and 
Berlin (in the latter he studied under Kirchhoff and Helmholtz), and 
received his doctorate from the University of Munich in 1879. Subsequently he 
taught in the universities of Munich (1880-1885), Kiel (1885-1889) and 
Berlin (1889-1928). In 1894 he became a member of the Prussian Academy of 
Science, and the permanent secretary of its Section of Mathematics and 
Natural Science in 1912. He was awarded the Nobel Prize for Physics in 1918. 
In 1930 he became president of the Kaiser Wilhelm Society in Berlin, later to 
be known as the Max Planck Society.

\section{References}

Bartoli, X., 1876, Fortschritte (2) 32, 888 and 1541.\\
Boltzmann, L., 1884, Annalen der Physik 22, 31.\\
Draper, J.W., 1847, Phil. Mag. 30, 345. \\
Herschel, W., 1800, Phil. Trans. Roy. Soc. London,
pages 255, 284, 293, 437 (also 1800, Phil. Mag. 7, 311 and {\em idem} 
8, pages 9, 16, 126, 253).\\
Kirchhoff, R., 1859, Monatsbericht d. Berl. Akad., December 1859, 783.\\
Kirchhoff, R., 1860, Annalen der Physik 109, 275.\\
Kurlbaum, F., 1898, Wied. Ann. 65, 759. \\
Leslie, J., 1804, {\em  Experimental Inquiry into the Nature and Propagation 
of Heat}.\\
Lummer, O. \& Pringsheim, E., 1899, Verh. d. deut. phys. Ges. 1, 23 and 215. \\
Lummer, O. \& Pringsheim, E., 1900, Verh. d. deut. phys. Ges. 2, 163. \\
Lummer, O. \& Pringsheim, E., 1901, Verh. d. deut. phys. Ges. 3, 36. \\
Maxwell, J. C., 1873, {\em A Treatise on Electricity and Magnetism}. \\
Paschen, F., 1899, Sitz. d. k. Akad. d. Wiss. zu Berlin, 405. \\
Planck, M., 1900Aa, Annalen der Physik 1, 69. \\
Planck, M., 1900Ab, Annalen der Physik 1, 719. \\
Planck, M., 1900Va, Verh. d. deut. phys. Ges. 2, 202. \\
Planck, M., 1900Vb, Verh. d. deut. phys. Ges. 2, 237. \\
Planck, M., 1901a, Annalen der Physik 4, 553. \\
Planck, M., 1901b, Annalen der Physik 6, 818. \\
Rayleigh, Lord (Strutt, J. W.), 1900, Phil. Mag. 49, 539.\\
Rubens, H., 1899, Wied. Ann. 69, 582. \\
Rubens, H. \& Kurlbaum, F., 1900, Sitz. d. k. Akad. d. Wiss. zu Berlin, 929. \\
Stefan, J.,1879, Sitz. d. k. Akad. d. Wiss. zu Wien, 79 (II), 391.\\
Thiesen, M., 1900, Verh. d. deut. phys. Ges. 2, 65.\\
Weber, H. F., 1887, Sitz. d. k. preuss. Akad. d. Wiss. zu Berlin, 491. \\
Wien, W., 1893, Sitz. d. k. preuss. Akad. zu Berlin, 55.\\
Wien, W., 1894, Annalen der Physik 52, 132.\\
Wien, W., 1896, Annalen der Physik 58, 662 (also 1897, Phil. Mag. 43, 214).\\

\end{document}